# Near 6 GHz Sezawa Mode Surface Acoustic Wave Resonators using AlScN on SiC

Xingyu Du[1], Nishant Sharma[1], Zichen Tang[1], Chloe Leblanc[1], Deep Jariwala[1], Roy H. Olsson III [1]

[1]*Department of Electrical and Systems Engineering, University of Pennsylvania, Philadelphia, PA, 19104, USA*

Surface Acoustic Wave (SAW) devices featuring Aluminum Scandium Nitride (AlScN) on a 4H-Silicon Carbide (SiC) substrate, offer a unique blend of high sound velocity, low thermal resistance, substantial piezoelectric response, simplified fabrication, as well as suitability for high-temperature and harsh environment operation. This study presents high-frequency SAW resonators employing AlScN thin films on SiC substrates, utilizing the second SAW mode (referred to as the Sezawa mode). The resonators achieve remarkable performance, boasting a $K^2$ value of 5.5% and a maximum Q-factor ($Q_{max}$) of 1048 at 4.7 GHz, outperforming previous benchmarks. Additionally, a SAW resonator with a 0.96 µm wavelength attains 5.9 GHz frequency with record $K^2$ (4.0%) and $Q_{max}$ (887). Our study underscores the potential of the AlScN on SiC platform for advanced radio-frequency applications.

---

a) Electronic mail: rolsson@seas.upenn.edu

Acoustic wave devices have established their dominance in high-volume, radio frequency (RF) front-end signal processing applications. Their superiority arises from impressive performance metrics, encompassing a compact size, minimal insertion loss, high isolation, low power consumption, precise filter responses, and potential for integration and co-fabrication alongside analog and RF electronics.[1] In order to meet the stringent requirements of upcoming wireless standards, the next generation of surface- and bulk-acoustic wave (SAW and BAW) resonators must exhibit an exceptionally sharp spectral response[2] which translates to both a high-quality factor (Q-factor) and a high electromechanical coupling coefficient ($K^2$). These attributes necessitate low propagation loss and a wider bandwidth, crucial for the performance of the acoustic filters.

SAW filters have been widely used in mobile phone handsets at frequencies up to 2 GHz due to their ease of fabrication, cost-effectiveness, and high repeatability.[3] However, achieving practical SAW devices operating at higher frequencies, such as 4 to 6 GHz, poses significant challenges, making it difficult for them to compete with their BAW counterparts.[2, 4] At these frequencies, the miniaturization of the interdigital transducers (IDTs) limits mass production, while other factors such as increased series resistance of the IDTs, reduced reliability due to poor heat dissipation and high thermal stress constrain their overall performance. Overcoming this frequency constraint involves leveraging high-velocity acoustic waves, leading to extensive research on acoustic wave devices operating above 3 GHz, making use of high-velocity SAW.[5-10]

Piezoelectric materials placed on silicon carbide (SiC) have previously demonstrated SAWs with significantly higher acoustic velocity compared to those on silicon, reaching values exceeding ~6600 m/s[11], 6310 m/s[10], and >10,000 m/s[12] for the second order SAW mode (referred to as the Sezawa mode).[10] This characteristic enables higher-frequency operation of SAW filters while effectively confining the acoustic wave energy near the SiC surface. SiC has also gained prominence in power electronics and high-temperature electronics due to its high thermal conductivity of 370 W/(m·K), which enhances the capability of SiC-based SAW filters to handle higher power levels before signal compression.[13] Additionally, SiC offers the advantage of a close lattice match with Aluminum Nitride (AlN) facilitating the growth of high-quality, crystalline c-axis oriented piezoelectric thin films.[11, 14-16] This feature simplifies the integration of piezoelectric materials with substrates at wafer scale.

Through alloying of Scandium (Sc) with Aluminum Nitride (AlN), previous studies have demonstrated a notable enhancement in the electromechanical coupling coefficient. The piezoelectric response coefficient for AlScN maximizes at around 42% Scandium concentration. Comparing $Al_{0.58}Sc_{0.42}N$ with pure AlN, the piezoelectric charge coefficient $d_{33}$ is 4 times and the transverse piezoelectric coefficient, $d_{31}$ is about 6 times higher.[17-20] In addition, high quality AlScN can be grown via sputtering at temperatures below 400 °C at lower costs and higher growth rates than epitaxial growth methods.[16]

This study presents SAW resonators using, ~1 um thick, $Al_{0.58}Sc_{0.42}N$ thin films on SiC substrates, achieving a remarkable $K^2$ up to 5.5% at 4.7 GHz and a $Q_{max}$ of 1048, surpassing previous works. The resonators exhibited a maximum Figure of Merit (FoM) of 38.4 at 4.9 GHz, doubling previous reports.[6, 10, 12, 21] Notably, a 0.96 μm wavelength SAW resonator device reached a frequency of 5.9 GHz while maintaining high $K^2$ of 4.0% and $Q_{max}$ of 887. We fabricate and compare various SAW resonator devices with different wavelengths in this study.

The fabrication of the AlScN devices adhered to the process flow depicted in Fig. 1(a). A 4-inch-high resistivity (> 1e5 ohm•cm) 4H-SiC substrates from Xiamen Powerway Advanced Material Co., Ltd were employed. Piezoelectric thin films were sputter deposited without breaking vacuum using an Evatec CLUSTERLINE 200 II PVD system at a temperature of 350 °C and a base pressure below 1 x $10^{-7}$ mbar. The deposition process consisted of three steps: a 15 nm AlN seed layer, a 35 nm gradient layer, and a 950 nm bulk AlScN layer, aimed at achieving optimal AlScN quality.[22] Throughout all three steps, a constant $N_2$ flow of 20 sccm was maintained, with no use of Ar as a process gas. For the 950 nm bulk AlScN layer with a 42% Sc concentration, the Al and Sc 4-inch target powers were set at 1 kW and 770 W, respectively. During the gradient layer deposition, the Sc target power was gradually increased from 0 to 770 W while keeping the Al target power constant. The chamber pressure remained approximately 8.0 x $10^{-4}$ mbar during deposition. After the AlScN film deposition, the films were subjected to analysis using Atomic Force Microscopy (AFM, Bruker Icon) and X-Ray Diffractometer (XRD, Rigaku Smart Lab), as depicted in Fig. 1 (b) and (c). Omega scan data of the deposited AlScN centered at 18.3° with a Full Width at Half Maximum (FWHM) of 1.43° indicating a high c-axis orientation of the film. This value was slightly greater than that observed in the $Al_{0.6}Sc_{0.4}N$ grown by molecular beam epitaxy, which had a value of 0.99°.[16] AFM scans revealed a root mean square roughness (Rq) of approximately 1.31 nm across a 10 x 10 µm² area, suggesting the absence of abnormally oriented grains (AOGs).[23]

Electron beam lithography (EBL) (Elionix ELS-7500EX) was used to pattern lines with various widths and spaces for interdigital transducers (IDTs). This was followed by Ti/Cu (10 nm/200 nm) ebeam evaporation and liftoff. Fig. 1 (d) provides an overview schematic of the one-port SAW resonators. Two Bragg mirrors positioned on both sides of the IDTs were employed as reflectors to confine the SAW. The wavelength of the SAW equaled twice the pitch of the IDTs and reflectors. Fig. 1 (e) displays Scanning Electron Microscope (SEM) images of the fabricated devices.

Resonator performance was measured using a Keysight vector network analyzer (VNA) P9374A with a power level of -20 dBm with 50 Ω port impedances. Prior to measurement, a two-port calibration was performed within the desired frequency range using the Short-Open-Load-Through (SOLT) method. Resonator parameters were manually extracted from the measured S-parameter data. Additionally, the widely recognized modified Butterworth Van Dyke (mBVD) circuit model allowed for further assessment of the fabricated resonators (Fig. 1 (f)).

From a resonator design perspective, understanding the relationship between the resonance frequency and the wavelength of the SAW is crucial, as it serves as a key fabrication parameter. Fig. 2 (a) shows the measured dispersion relationships of the Sezawa and Rayleigh modes of the SAW devices. The devices utilize an aperture width of 40 µm, 125 IDT finger pairs, and 240 reflectors unless otherwise stated. The Rayleigh mode represents the fundamental mode with a frequency range spanning from 1 to 4 GHz. The Sezawa mode is the second mode at even higher frequencies, typically displaying higher $K^2$ and Q-factors when the piezoelectric layer thickness and IDT design remain constant.[10, 24] Due to its substantial phase velocity, the Sezawa mode is utilized for high-frequency SAW resonator fabrication and demonstration. For simulations, COMSOL Multiphysics® version 5.6 was employed to model the SAW resonators featuring a 10 nm Ti layer and a 200 nm Cu layer atop a 1 μm AlScN on a SiC substrate. In this finite element method (FEM) simulation, an infinitely repeated IDT finger pairs assumption was applied, allowing for the simulation of just one pair of IDT fingers with periodic repetition. The distinction between the Sezawa and Rayleigh modes becomes evident in the total mechanical displacement field at the resonance frequencies corresponding to the Rayleigh and Sezawa modes, as depicted in Fig. 2 (b) and (c). Specifically, the Rayleigh mode is localized near the top surface (AlScN) of the SAW resonator, whereas the Sezawa mode is confined to the interface between AlScN and SiC.

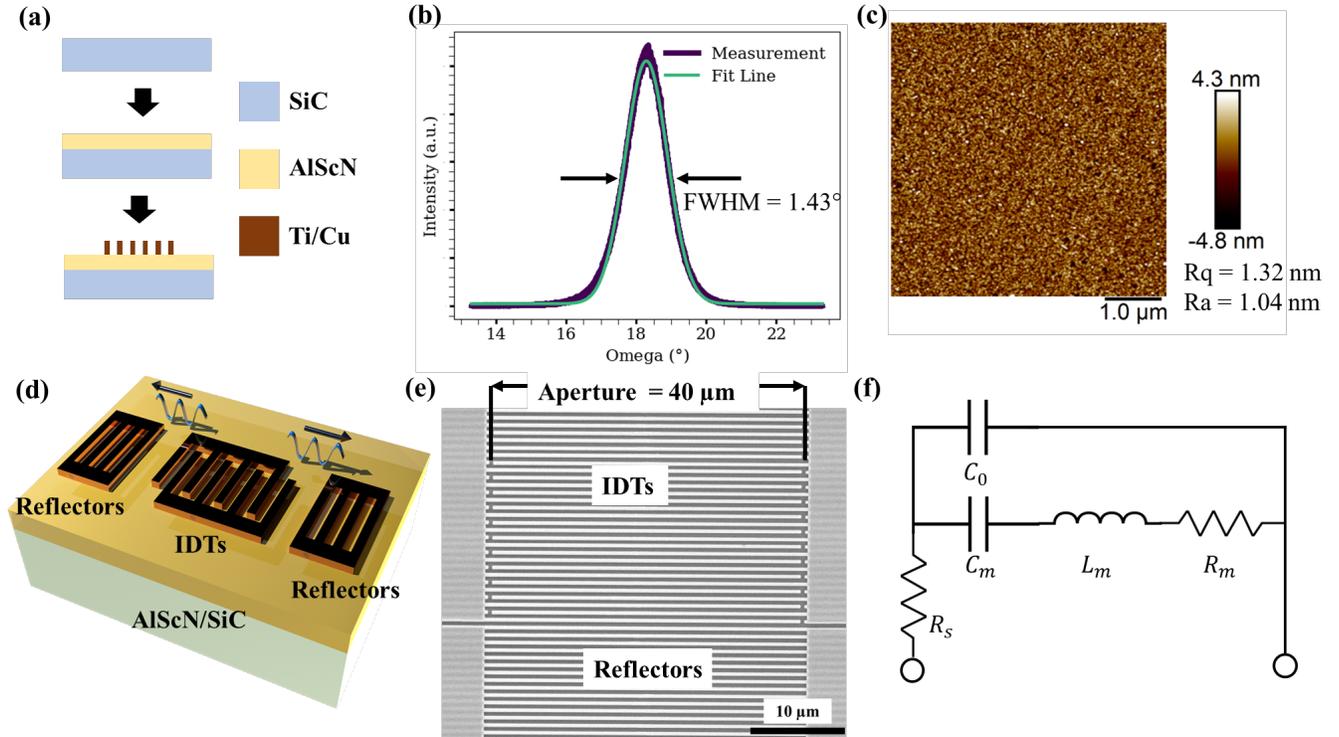

**FIG. 1**. (a) Process flow for fabrication of the AlScN/SiC SAW resonators (b) XRD rocking curve measurement of 1 μm AlScN on top of SiC. The Full Width at Half Maximum (FWHM) is 1.43°. (c) AFM image of 1 μm thick AlScN film deposited on SiC. Root mean square roughness (Rq) and average roughness (Ra) are 1.32 nm and 1.04 nm, respectively. (d) Schematic of a one-port SAW resonator (e) Scanning electron microscope (SEM) image focusing on the IDTs and Reflectors. (f) mBVD equivalent circuit model of the SAW Resonator.

Fig. 2 (d) and (e) showcase the measured admittance response of the SAW resonators within the frequency range of 1 to 7 GHz. Fig. 2 (d) shows the measured admittances of the Rayleigh modes while Fig. 2 (e) displays the frequency response of the Sezawa modes. The resonance frequencies of both the Rayleigh modes and Sezawa modes increase as the designed wavelength of the IDT decreases, with minimal spurs observed outside the Rayleigh and Sezawa frequency band.

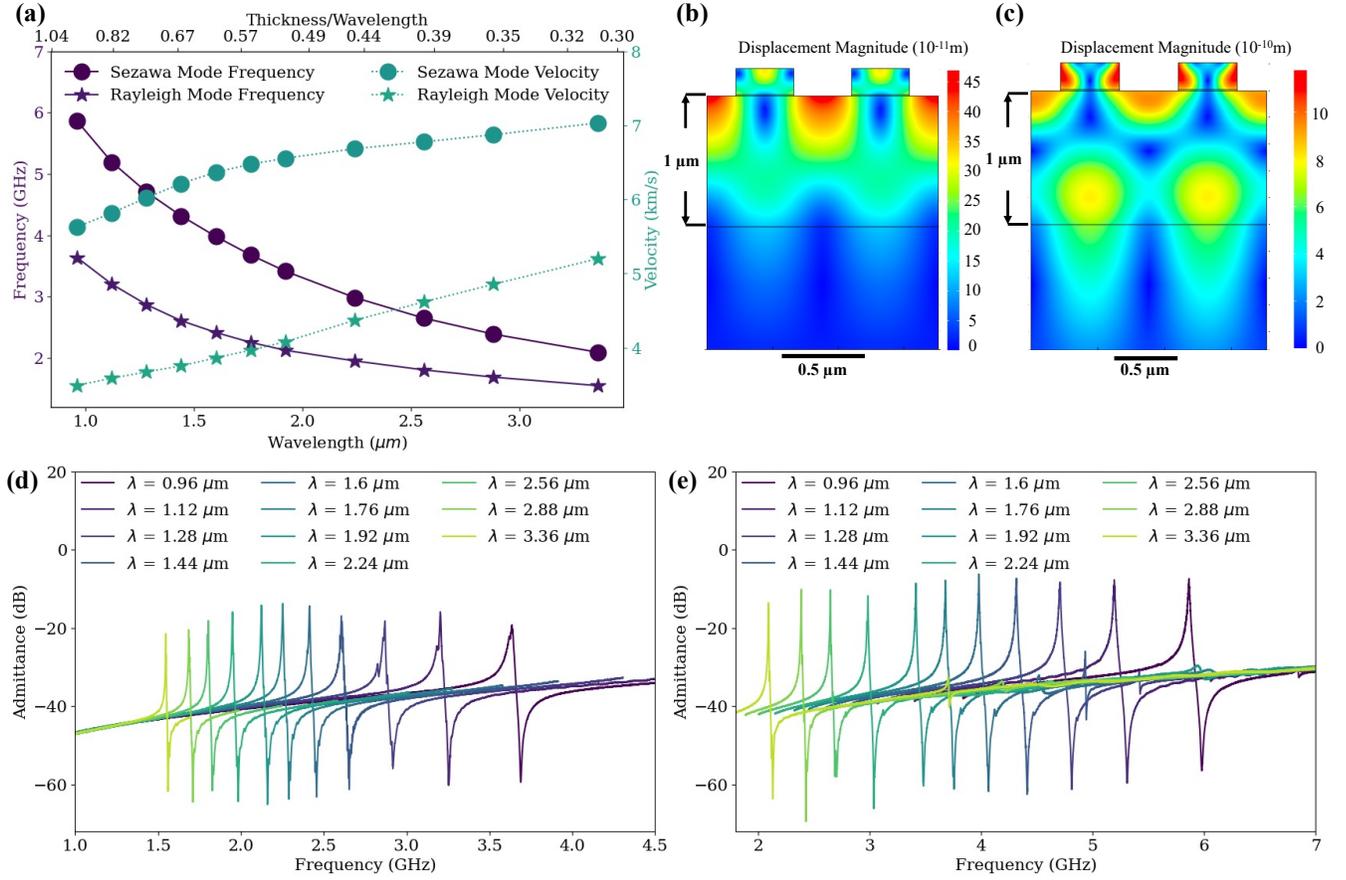

**FIG. 2**. (a) Measurement results showcasing the resonance frequencies ($f_s$) of the SAW resonators' Rayleigh modes and Sezawa modes. The dashed and solid lines are a guide to the eye. (b-c) Finite Element Method (FEM) simulations displaying the mode displacement of a 1 μm AlScN-on-SiC device equipped with 10 nm Ti and 200 nm Cu electrodes for (b) the Rayleigh mode and (c) the Sezawa mode. (d) Admittance curves depicting the measured Rayleigh modes of the fabricated SAW resonators with varying wavelengths. (e) Admittance curves illustrating the measured Sezawa modes of the fabricated SAW resonators with differing wavelengths.

Fig.3 compares the relationship of $K^2$ and Q-factor of both the Sezawa and Rayleigh modes of the SAW resonators vs. wavelength and frequency. The Q-factor is defined as the frequency of the series resonance ($f_s$) or parallel resonance ($f_p$) over its 3dB-bandwidth and the $K^2$ is calculated using the formula:

$$K^2 = \frac{f_p^2 - f_s^2}{f_s^2} \qquad (1)$$

To account for the influence of spurs between $f_s$ and $f_p$, which can cause inaccurate estimation of the Q-factor and $K^2$, the measured admittance responses were fit using the mBVD circuit model and $K^2$, $Q_s$, and $Q_p$ were calculated accordingly. The Sezawa mode's $K^2$ reaches a maximum of 4.6% at 4.3 GHz at a wavelength of 1.44 μm, corresponding to an AlScN thickness/wavelength ratio (t/λ) of 0.69. Within the wavelength range of 1.28 μm and 2.4 μm where the frequencies range from 1.9 GHz to 2.9 GHz, the $K^2$ maintains a high value of approximately 3.5%. These reported measured $K^2$ values align closely with the maximum values obtained from previous theoretical calculations. Hashimoto et al. reported a simulated maximum of 5.26% at t/λ=0.58 for $Al_{0.58}Sc_{0.42}N/SiC$[10], while Gokhale et al. reported a simulated maximum of 2.5% at t/λ=0.35 for $Al_{0.75}Sc_{0.25}N/SiC$.[12] It is important to note that the absence of consideration for metal electrodes in these simulations could account for the difference between the simulated and experimental results, as copper electrodes have the potential to store strain energy and reduce the coupling coefficient.

In the Sezawa mode, the Q-factor shows an increasing trend with wavelength, reaching a maximum $Q_s$ of 893 and a maximum $Q_p$ of 1752 at a frequency of 2.6 GHz. This phenomenon might be attributed to changes in copper reflectivity with the SAW wavelength. The copper reflector aids in confining energy at larger wavelengths, resulting in higher Q-factors. For the Rayleigh mode, the Q-factor does not change significantly with wavelength. Both the Sezawa and Rayleigh modes have significantly higher $Q_p$ values compared to $Q_s$. These devices typically have a series resistance of around 1~2 Ω that can be reduced, for instance, by decreasing the SAW aperture to enhance the overall $Q_s$.

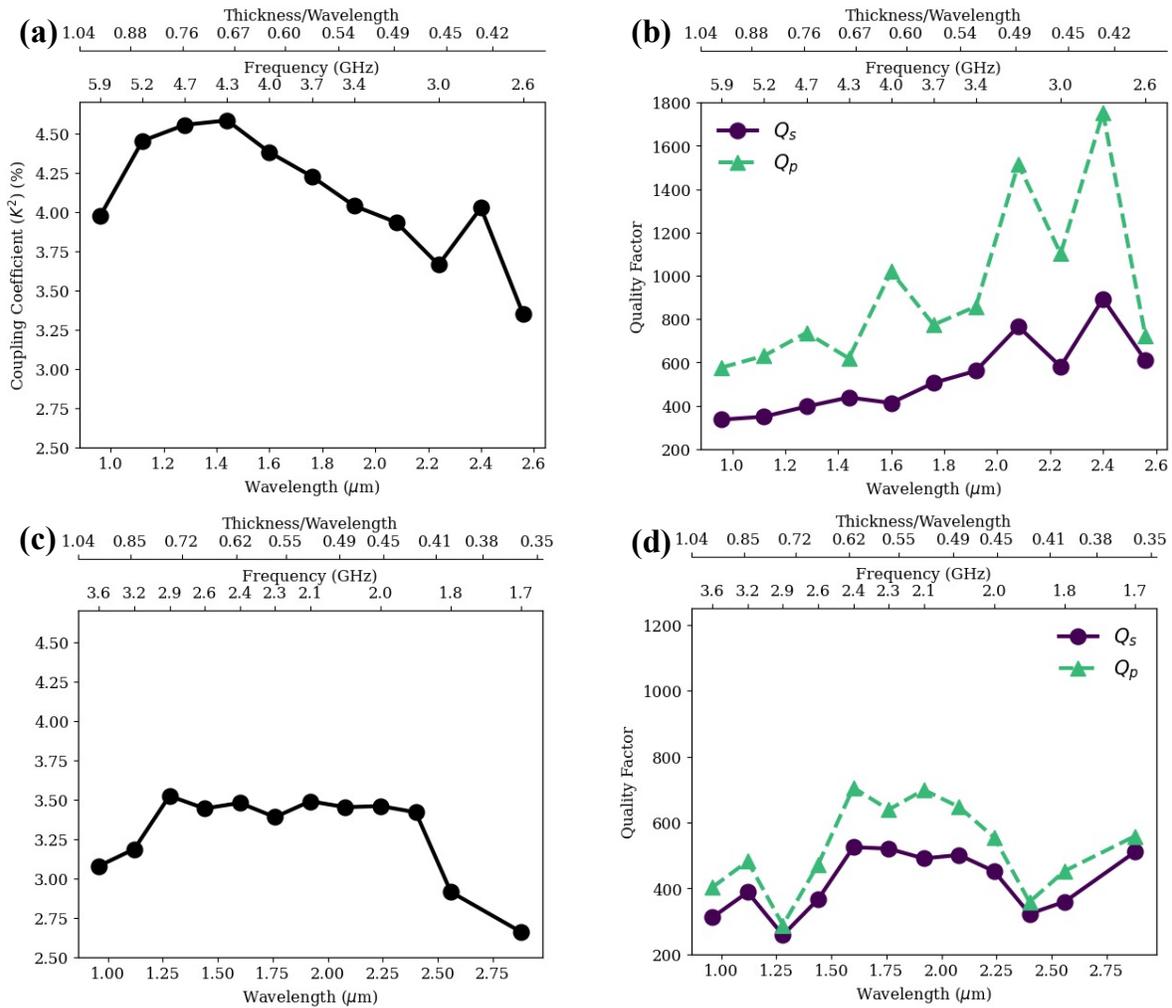

**FIG. 3**. (a) Measured Sezawa mode K$^2$ vs. wavelength or frequency (b) Measured Sezawa mode Quality factor (Q-factor) vs. wavelength or frequency (c) Measured Rayleigh mode K$^2$ vs. wavelength or frequency (b) Measured Rayleigh mode Q-factor vs. wavelength or frequency

In Fig. 1 of the Supporting Information (SI), the scenario is depicted where spurs overlap with f$_p$, rendering the direct 3dB bandwidth Q-factor calculation from the measured response invalid. This undesirable effect is attributed to the presence of a spurious mode known as the transverse mode. For devices with frequencies above 4 GHz, the spurs are nearly negligible due to the high velocity, causing the transverse mode frequencies to fall outside the Sezawa frequency band. Several techniques have been introduced previously to suppress these transverse modes, such as apodization IDT[25], piston mode IDT[26], and tilted IDT[27].

SAW resonators featuring varying numbers of IDT finger pairs were fabricated and compared in Fig. 4. These devices were designed with an aperture width of 40 μm and 200 reflectors, corresponding to a wavelength of 1.6 μm and a resonance frequency of approximately 4.3 GHz. As the number of IDT finger pairs increased, the admittance response shifted upward to smaller impedances due to the rise in parallel capacitance ($C_0$). With an increase in the number of IDT finger pairs to 550, the series resonance peak broadened, leading to a decrease in both $Q_s$ and the motional impedance $R_m$. However, $K^2$ increased with the number of IDT finger pairs, indicating that more electrical energy coupled into mechanical energy as the coupling region expanded. In SI Fig. 2, more IDT finger pairs were added to enhance $K^2$, resulting in an increase from 4.6% to 5.5%. However, this improvement came at the cost of reduced $Q_s$, which decreased from 396 to 164 when compared to other 4.7 GHz devices. It can be concluded that a significant amount of energy is stored in the reflectors, degrading the $K^2$. Future studies utilizing reflector materials with higher reflectivity than Copper could maintain the same Q-factor while improving $K^2$ by better confining the SAW energy.

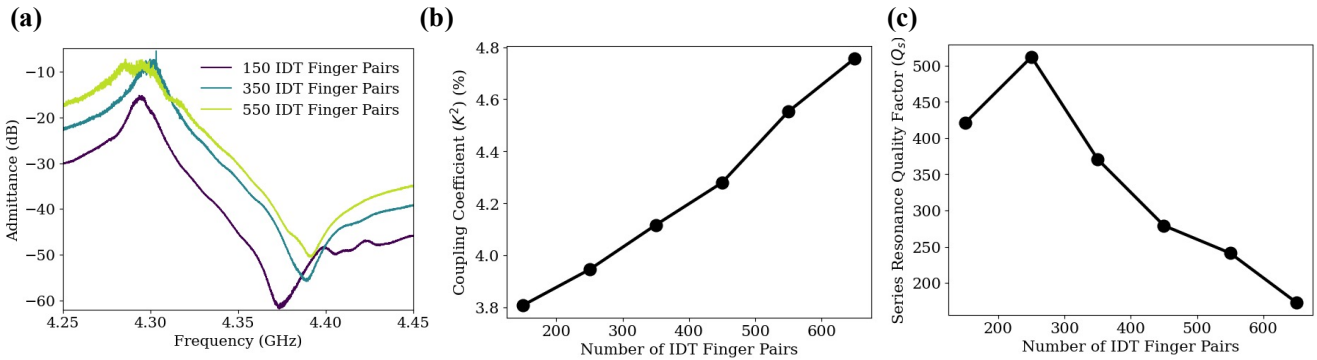

**FIG. 4**. (a) Comparative analysis of admittance responses based on varying numbers of IDT finger pairs. (b) Comparison of the $K^2$ values concerning different number of IDT finger pairs. (c) A comparison of the $Q_s$ with varying numbers of IDT finger pairs.

Fig.5 shows the measured response of four different SAW resonators with frequencies ranging from 4~6 GHz, demonstrating the high $K^2$ and Q-factor of the fabricated devices. The devices were designed with an aperture width of 40 μm, 125 IDT finger pairs, and 240 reflectors and with different wavelengths from 0.96 μm and 1.44 μm. The circuit model showed a good agreement with the measured frequency response. All four resonators exhibit an $R_s$ of around 1 Ω and $C_0$ of around 0.7 pF.

Another method to evaluate Q-factor is using the Bode-Q ($Q_{max}$), which is defined by using the phase group delay of the S-parameter. In order to calculate $Q_{max}$ of the SAW resonators, all of the S-parameter data must lie nearly equidistant from the center of the Smith chart.[28, 29] As shown in the center plot in Fig. 5, a 50 Ω source impedance is not sufficient to center the smith circuit. SI Fig. 3 shows the continuous Q circle placed in the center of the Smith chart using source impedance matching.

Table 1 provides a comparison of the SAW resonators with state-of-the-art high-frequency SAW counterparts from the literature. The Figure of Merit (FoM) defined as the $K^2 \times Q_p$, reaches a peak value of 38.4 at a frequency of 4.7 GHz in the resonator devices. This FoM value is more than double that of previous work utilizing AlScN. In this study, a maximum resonance frequency of 5.9 GHz was achieved, surpassing the previously highest frequency of 4.61 GHz achieved for AlScN based SAW resonators. A recent study demonstrated a leaky SAW resonator using 42° X-cut LiNbO$_3$ on SiC achieved a remarkably high $k_t^2$ of up to 19.6 % at 4.95 GHz[5]. Despite confining the SAW energy near the substrate surface, its leakage to the substrate remains non-negligible, limiting the Q-factor. It is noteworthy that AlScN/SiC based SAW resonators offer several advantages over LiNbO$_3$/SiC based SAW resonators. In this study, AlScN could be directly sputtered on top of SiC, simplifying the fabrication process and reducing costs. In contrast, LiNbO$_3$ thin films require complex steps such as ion implantation, exfoliation, transferring, post-annealing, and chemical-mechanical polishing for integration with SiC[5, 30]. Additionally, AlScN/SiC based SAW devices are suitable for applications requiring high temperatures or integration with processes requiring elevated temperatures. The high pyroelectric coefficient of LiNbO3 renders it vulnerable to shattering under rapid temperature changes.[31, 32]

In conclusion, we successfully demonstrate high-performance SAW resonators utilizing a sputtered $Al_{0.58}Sc_{0.42}N$-on-SiC platform. The designed structure of the SAW resonator has been optimized to simultaneously achieve high frequency, high $K^2$, high Q-factor, and weak spurious modes. We attribute this to the exceptional piezoelectric properties of the AlScN thin films and the lattice matching of the SiC substrate. The impact of varying IDT finger pairs on performance was thoroughly explored. The frequency response trends concerning different wavelengths and frequencies were also outlined. While further optimization of IDT finger geometry is required to address spurious modes in the low-frequency SAW resonators, this study underscores the promising potential of AlScN/SiC acoustic devices for radio-frequency applications.

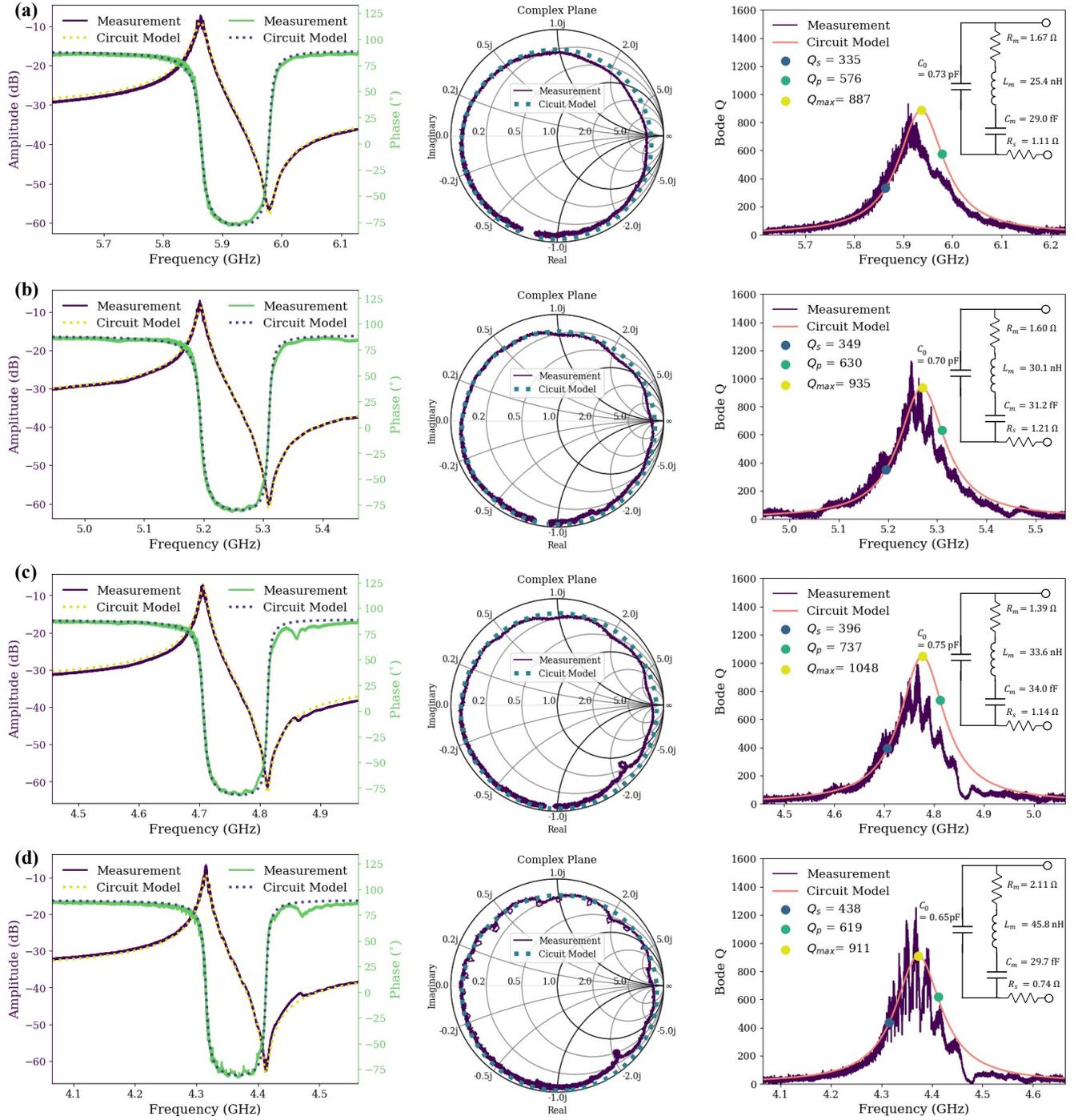

**FIG. 5**. Admittance response (left), Smith plot with 50 Ω reference impedance (center), $Q_{max}$ with changed impedance (right) of Sezawa mode SAW devices with wavelength of (a) 0.96 μm, (b) 1.12 μm, (c) 1.28 μm, and (d) 1.44 μm.

Table 1 Comparison of high frequency AlScN surface acoustic wave resonators

| Year | Material | Freq (GHz) | $K^2$ | Qs | Qp | FoM | $Q_{max}$ |
|---|---|---|---|---|---|---|---|
| This work | $Al_{0.58}Sc_{0.42}N$ /SiC | 5.9 | 4.0% | 335 | 576 | 23.0 | 887 |
| | | 5.2 | 4.5% | 349 | 630 | 28.4 | 935 |
| | | 4.7 | 4.6% | 396 | 737 | 33.9 | 1048 |
| | | 4.7 | 5.5% | 164 | 698 | 38.4 | 893 |
| | | 4.3 | 4.6% | 438 | 619 | 28.5 | 911 |
| 2023[5] | X cut $LiNbO_3$/SiC | 4.95 | 15.9% | - | - | - | 408 |
| 2023[6] | $Al_{0.66}Sc_{0.34}N$/Diamond | 3.71 | 6.3% | 120 | 23 | 7.6 | - |
| | $Al_{0.57}Sc_{0.43}N$/Diamond | 3.73 | 5.4% | 311 | 22 | 16.6 | - |
| 2022[33] | $Al_{0.8}Sc_{0.2}N$/GaN/Sapphire | 4.61 | - | - | - | - | 571 |
| 2020[12] | $Al_{0.75}Sc_{0.25}N$/SiC | 3.73 | 0.5% | - | 108 | 0.6 | |
| 2019[21] | $Al_{0.88}Sc_{0.12}N$/Si | 3.6 | 3.0% | - | 146 | 5.4 | - |
| 2013[10] | $Al_{0.6}Sc_{0.4}N$/SiC | 3.8 | 4.5% | 340 | 240 | 15.3 | - |

This work was funded in part by the DARPA Traveling-Wave Energy Enhancement Devices (TWEED) program under award HR0011-21-9-0055. This work was also funded in part by the NSF CAREER Award (1944248). This work was carried out in part at the Singh Center for Nanotechnology at the University of Pennsylvania, a member of the National Nanotechnology Coordinated Infrastructure (NNCI) network, which is supported by the National Science Foundation (Grant No. HR0011-21-9-0004).

**AUTHOR DECLARATIONS**
**Conflict of Interest**
The authors have no conflicts to disclose.

**Author Contributions**
**Xingyu Du:** Conceptualization (lead); Investigation (lead); Methodology (lead); Writing – original draft (lead); Writing – review & editing (equal). **Nishant Sharma:** Investigation (supporting). **Zichen Tang:** Investigation (supporting); Methodology (supporting). **Chloe Leblanc:** Investigation (supporting). **Deep Jariwala:** Supervision (lead); Funding acquisition (lead). **Roy H Olsson III:** Conceptualization (lead); Supervision (lead); Writing – review & editing (lead); Funding acquisition (lead)

**DATA AVAILABILITY**
The data that support the findings of this study are available from the corresponding author upon reasonable request.

## Support Information

## Near 6 GHz Sezawa Mode Surface Acoustic Wave Resonators using AlScN on SiC

Xingyu Du[1], Nishant Sharma[1], Zichen Tang[1], Chloe Leblanc[1], Deep Jariwala[1], Roy H Olsson III [1]

[1]*Department of Electrical and Systems Engineering, University of Pennsylvania, Philadelphia, PA, 19104, USA*

In the Supporting Information (SI) Fig. 1 (a), one device exhibits severe spurs between $f_s$ and $f_p$ in the admittance response. These spurs are spaced approximately 7 MHz apart in frequency and appear as circles on the Smith chart, representing other resonance modes. SI Fig. 1 (b) illustrates a scenario where these spurs overlap with $f_p$, invalidating the direct 3dB bandwidth Q-factor calculation from the measured response. This spurious mode is known as the transverse mode. Although both devices operate with similar frequency, the Sezawa mode depicted in SI Fig. 1 (b) operates at a much higher velocity than SI Fig. 1(a), resulting in a larger frequency spacing.

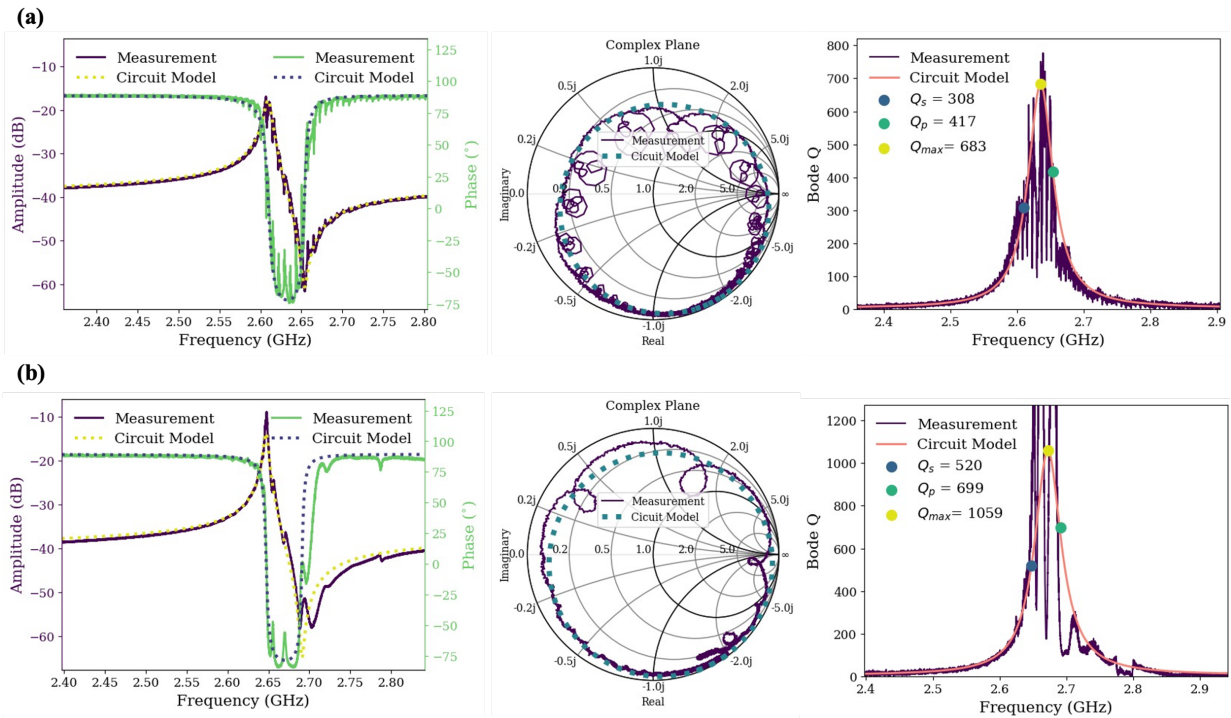

**FIG. 1**. Admittance response (left), Smith plot with 50 Ω reference impedance (center), $Q_{max}$ with changed source impedance for (a) Sezawa mode with wavelength of 2.56 μm and (b) Sezawa mode with wavelength of 1.44 μm.

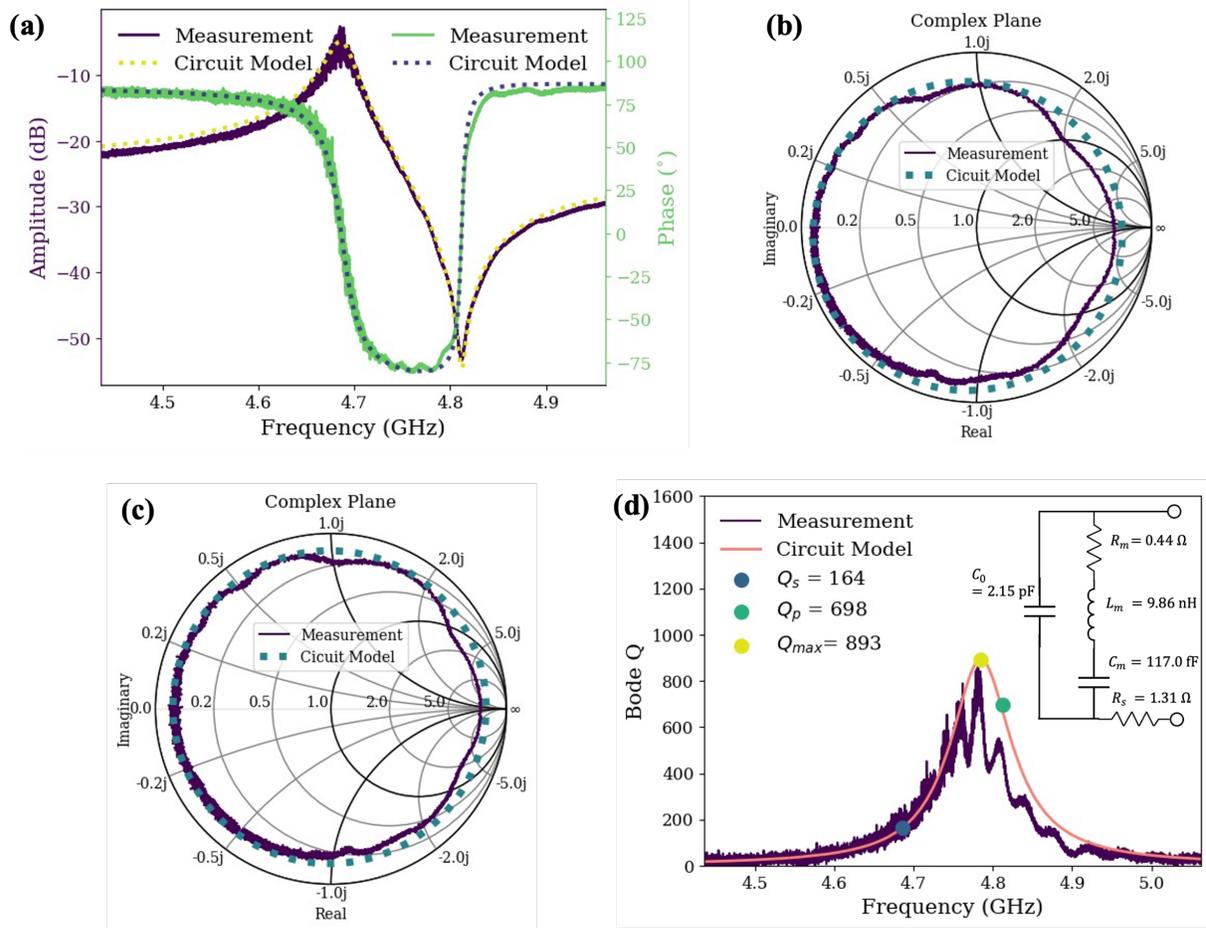

**FIG. 2**. Sezawa mode SAW resonator showing $K^2$ of 5.5% with wavelength of 1.44 μm, aperture of 44 μm, 445 IDT finger pairs, and 240 reflectors: (a) Admittance response, (b) Smith plot with 50 Ω reference impedance, (c) Smith plot with 33-2.2j Ω reference impedance, (d) $Q_{Bode}$ with 33-2.2j Ω reference impedance.

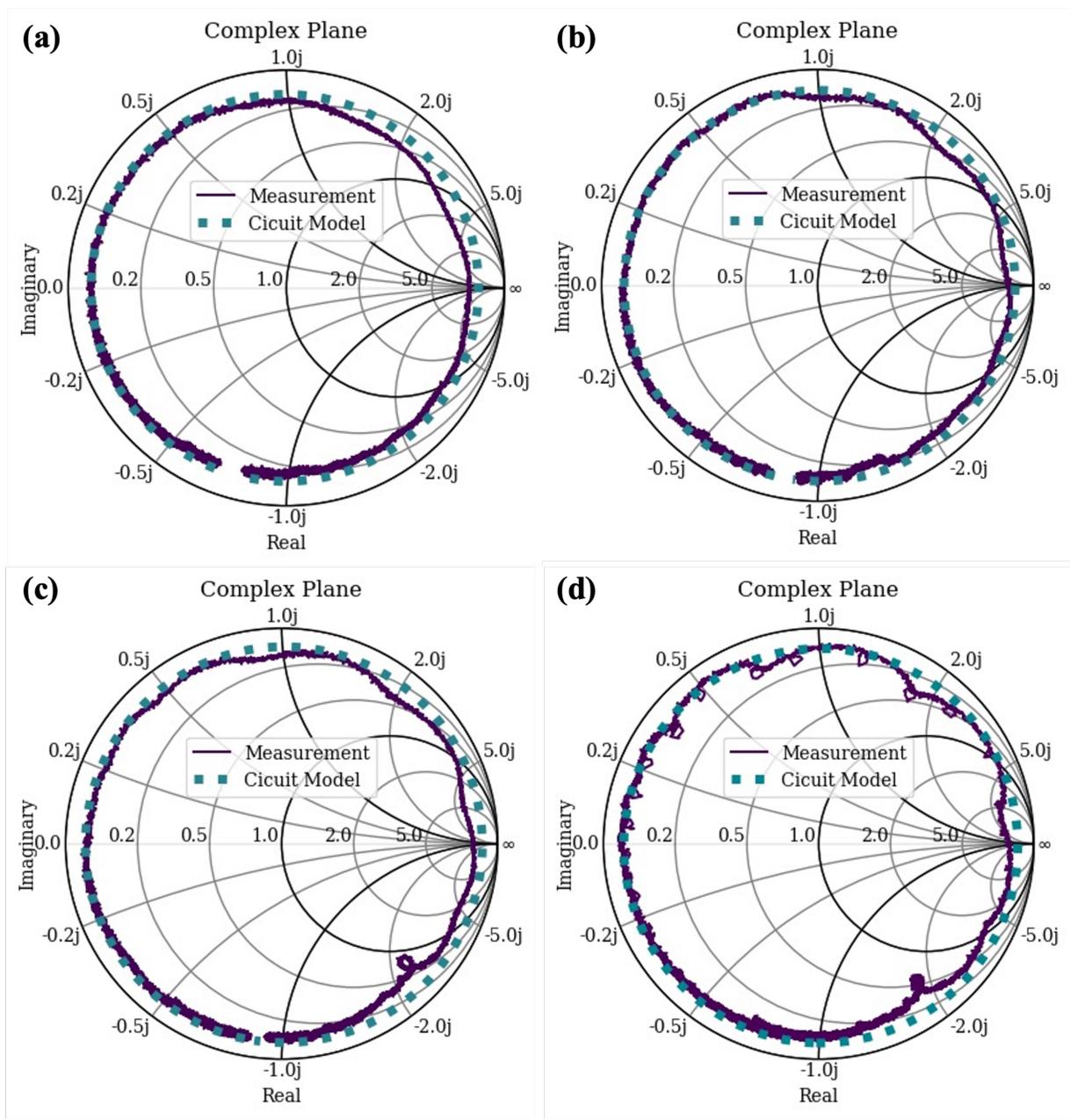

**FIG. 3**. The measured and circuit model results plotted on a Smith chart after transforming the source impedance of (a) 50-4.5j Ω for a wavelength of 0.96 μm, (b) 51-3.6j Ω for a wavelength of 1.12 μm, (c) 52-2.9j Ω for a wavelength of 1.28 μm, and (d) 57-3.7j Ω for a wavelength of 1.44 μm.